\documentclass[prb,twocolumn,showpacs,amsmath,amssymb,superscriptaddress]{revtex4}
\usepackage{graphicx}
\usepackage{graphics}
\usepackage{dcolumn}
\usepackage{bm}
\usepackage{amsmath,amssymb,latexsym}

\newcommand{\rr}{{\bf r}}
\newcommand{\kk}{{\bf k}}
\newcommand{\MM}{{\bf M}}
\newcommand{\BB}{{\bf B}}

\begin{document}
\title{Magnetization relaxation in (Ga,Mn)As ferromagnetic semiconductors}
\author{Jairo Sinova}
\affiliation{Physics Department, Texas A\&M University, College Station, TX 77843-4242}
\author{T. Jungwirth}
\affiliation{Institute of Physics  ASCR, Cukrovarnick\'a 10, 162 53  
Praha 6, Czech Republic }
\affiliation{Physics Department, University of Texas at Austin, Austin TX 78712-0264} 
\author{X. Liu}
\affiliation{Department of Physics, University of Notre Dame,
Notre Dame, IN 46556}
\author{Y. Sasaki}
\affiliation{Department of Physics, University of Notre Dame,
Notre Dame, IN 46556}
\author{J.K. Furdyna}
\affiliation{Department of Physics, University of Notre Dame,
Notre Dame, IN 46556}
\affiliation{Physics Department, University of Texas at Austin, Austin TX 78712-0264}  
\author{W. A.  Atkinson}
\affiliation{Physics Department, Trent University, Ontario, Canada K9J 7B8}
\author{A.H. MacDonald} 
\affiliation{Physics Department, University of Texas at Austin, Austin TX 78712-0264}
\date{\today}
\begin{abstract}

We describe  a theory of Mn local-moment  magnetization relaxation due
to p-d kinetic-exchange coupling  with the itinerant-spin subsystem in
the ferromagnetic semiconductor (Ga,Mn)As alloy.  The theoretical
Gilbert damping coefficient implied  by this  mechanism is
calculated as a function of Mn moment density, hole concentration, and
quasiparticle lifetime. Comparison with experimental ferromagnetic
resonance data suggests that in annealed strongly metallic samples, 
p-d coupling contributes significantly to the damping rate 
of the magnetization precession 
at low temperatures. By combining the theoretical Gilbert coefficient
with the values of the magnetic anisotropy energy, 
we estimate that the typical 
critical current for spin-transfer magnetization switching in all-semiconductor
trilayer devices can be as low as
$\sim 10^{5} {\rm A \;cm}^{-2}$.

\end{abstract}

\maketitle 

\section{Introduction}  
The Gilbert coefficient describes  the damping of
small-angle magnetization precession and  is one of the key parameters
that characterizes collective magnetization dynamics in a ferromagnet.
Early theories  of magnetization dynamics in  transition metals viewed
exchange  coupling  ($\propto  {\bf  S}\cdot{\bf s}$)  between  local
moment d-shell spins ${\bf S}$  and itinerant s-p band 
spins ${\bf s}$  as  a key relaxation mechanism.\cite{kittel_pr56}   
We now recognize
that  this model  needs  to  be elaborated  for  transition metals  to
account for  the itinerant character of their  d-electrons.  Models of
d-shell local moments that  are exchange-coupled to itinerant s-p band
electrons have, however, been  resurrected recently because they provide
a  good  description  of   ferromagnetism  in  many  diluted  magnetic
semiconductors (DMSs),         (Ga,Mn)As         in         
particular.\cite{dietl_handbook94,ohno_jmmm99}   
Exchange-coupling between local
moments and  itinerant electrons should  also contribute significantly
to Gilbert damping in these new ferromagnetic systems.  The elementary
process  for this  damping mechanism  is one  in which  a local-moment
magnon  is  annihilated  by  exchange interaction  with  a  band
electron  that suffers  a spin-flip.   This process  cannot  by itself
change  the  total  magnetic  moment since  the  exchange  Hamiltonian
commutes with the total spin ${\bf S}+{\bf s}$.  Net relaxation of the
magnetization  requires  another  independent  process  in  which  the
itinerant electron spin relaxes through spin-orbit interactions.

Recent  experiments   and  ab-initio  calculations   \cite{mm03}  have
established  that  for  doping  levels  up  to  several  per  cent,  a
substitutional Mn impurity in GaAs introduces five strongly localized
d-electrons  and  a  delocalized  hole  in the  As/Ga  p-band,  and  that
ferromagnetic coupling  between the $S=5/2$ Mn moments  is mediated by
the p-d kinetic-exchange. Hence,  we model \cite{bookchapter}
the electronic structure of
the free carriers by that  of the host material, implicitly assuming a
shallow  acceptor picture.  The  free carrier  quasiparticles are p-d 
exchange coupled to the local moments with strength \cite{ohno_jmmm99} 
$J_{pd}\approx 55$~meV~nm$^3$, and
have a finite life time that can be estimated perturbatively. 
This  theoretical picture leads
to  an  accurate  description  of  many  thermodynamic  and  transport
properties  of  optimally  annealed  (Ga,Mn)As samples,  such  as  the
measured  transition temperature,\cite{dietl_sci00,tc}  the anomalous
Hall    effect,\cite{ohno_jmmm99,jungwirth_0302060}    anisotropic
magnetoresistance,\cite{jungwirth_0302060}   and   magneto-optical
properties. \cite{dietl_sci00,sinova_prb02}  Particularly important in
justifying the present theory  are results for the magneto-crystalline
anisotropy,\cite{aniso}  spin-stiffness, \cite{konig_prb01,stiffness}
and  Bloch  domain  width  \cite{domains}  that all  agree  well  with
experiment.  These parameters follow from the long-wavelength limit of
the theory  of the Mn  spin-wave dispersion, and reflect  the retarded
and non-local character of the valence-band-hole mediated interactions
between  Mn  moments.\cite{konig_prb01}   The  Gilbert  damping  of
magnetization  precession  discussed  here  is  the  aspect  of  this
long-wavelength  collective   magnetization  dynamics  that   is  most
directly dependent on valence-band spin-orbit coupling.

In Section II of this paper, we  present a fully microscopic theory of
the  kinetic-exchange  contribution  to  the  Gilbert  coefficient  in
DMSs.  By  comparing the linear  response  predicted  by  the  classical
phenomenological Landau-Lifshitz-Gilbert   (LLG)    equation   with
microscopic   linear  response   theory,  we   identify   the  Gilbert
coefficient with the dissipative part of a susceptibility diagram.

In Section III we present experimental ferromagnetic
resonance (FMR) data \cite{furdyna,stiffness} 
recorded as a function of temperature and external
magnetic field strength and orientation. Since the frequency dependence
of the FMR linewidth is not available,\cite{furdyna,stiffness}
we are unable to experimentally
decouple inhomogeneous FMR broadening and the
{\em intrinsic} Gilbert damping contributions
to the linewidth to make a quantitative comparison with the theory.
Nevertheless, the data indicate that the magnetic inhomogeneity
contribution is largely suppressed in the more metallic, annealed
samples and that much of the observed low-temperature FMR linewidth in 
these samples
can be explained by  damping of the magnetization precession mediated 
through the p-d coupling.

By adding  a spin-torque
term  \cite{slonczewski_jmmm96} to the  LLG equation,  we 
estimate  in Section IV 
that the typical critical  current for  magnetization switching  due to
spin-transfer torques in an all-semiconductor trilayer device consisting
of  magnetically  ``soft'' and  ``hard''  DMS  layers  separated by  a
non-magnetic  spacer will  be $\sim  10^{5} {\rm  A \;  cm}^{-2}$. 
In
metals,  this  spin-transfer  effect  is  currently  the  focus  of  a
considerable experimental \cite{stexp} and theoretical \cite{sttheory}
research.
Spin-transfer switching has  not yet been  demonstrated in
all-semiconductor  systems, but the  effect promises to have  a richer
phenomenology in this case because of the flexibility of semiconductor
ferromagnet  materials, and  because  of the  possibility of  combining
spin-transfer with other semiconductor circuit 
functionalities.\cite{field_tc,field_reversal} The relatively
low critical currents we predict for semiconductors may also circumvent the
incomplete magnetization switching encountered in metallic
magnetic tunnel junctions that occurs
due to the interference of strong self-field
effects with the spin-transfer torques.\cite{vortex}

The paper concludes with a brief summary of our results.

\section{Theory of the Gilbert damping}
{\em Semiclassical LLG Linear Response.}
The phenomenological LLG equation for collective magnetization dynamics is 
\begin{equation}
\frac{d\MM}{dt}=-\frac{g\mu_B}{\hbar}\MM\times\BB_{eff}
+\frac{\alpha}{M}\MM\times\frac{d\MM}{dt}\; ,
\label{llg}
\end{equation}
where $\MM$ is the local Mn moment magnetization, $\BB_{eff}=-\partial
E/\partial\MM$  is the effective  magnetic field,  $g$ is  the Land\'e
g-factor,  $\mu_B$   is  the  Bohr  magneton,  and   $\alpha$  is  the
phenomenological   Gilbert   damping   coefficient.  Unless $\alpha$ 
depends strongly on the orientation of the magnetization \cite{rem_alpha_angle} or if
the magnetization is not fully aligned with the external static magnetic field,
the Gilbert damping rate, observed in experiment through a frequency-dependent
FMR linewidth, is independent of the static field  and of the details
of magnetic anisotropies present in the sample.\cite{vonsovskii} This allows us to
 assume in this section a simple geometry in which  the anisotropy fields are represented
by a single, uniaxial anisotropy energy density coefficient $U$. For small
fluctuations of the Mn magnetization orientation around the easy axis,
Eq.~(\ref{llg})  can be  used  to derive  a phenomenological  response
function  of the  magnetic  system  to a  weak  transverse field.  For
zero external static magnetic field the
corresponding inverse susceptibility reads:
\begin{equation}
\chi^{-1}=\frac{\hbar}{(g\mu_B)^2N_{Mn}S}
\left(\begin{array}{cc}
\tilde{U}-i\alpha\omega & -i\omega\\
i\omega & \tilde{U}-i\alpha\omega\\
\end{array}\right)\; ,
\label{classicalsusc}
\end{equation} 
where  $\tilde{U}=U/(\hbar N_{Mn}S)$,   
$N_{Mn}=4x/a^3_{lc}$
is the  density of uniformly  distributed Mn
moments in Ga$_{1-x}$Mn$_x$As ($a_{lc}$ is the GaAs lattice constant), 
and $\omega$ is the  frequency of the external rf field
perturbation.                

{\em Microscopic theory.}
We  derive the zero-temperature quantum response function from our 
effective Hamiltonian theory and obtain a microscopic expression
for $\alpha$ by equating the quantum mechanical response function to the
classical one in the uniform $\omega\rightarrow 0$ limit.
We start by writing a quantum analog of Eq.~(\ref{llg}) 
using the linear response theory,
\begin{eqnarray}
& &\langle M_x (\rr,t)\rangle=-\frac{i}{\hbar}\int_{-\infty}^{\infty}
dt^{\prime}\int d\rr^{\prime}
\bigg(\langle[M_x (\rr,t),-M_x (\rr^{\prime},t^{\prime})
\nonumber \\
& &B_x(t^{\prime})]
\rangle+
\langle[M_x (\rr,t),-M_y (\rr^{\prime},t^{\prime})B_y(t^{\prime})]
\rangle\bigg)\theta(t-t^{\prime})\; ,
\label{lrt}
\end{eqnarray}    
which leads to the  retarded transverse susceptibility:
\begin{equation}
\chi^R_{i,j}(\rr,t|\rr^{\prime},t^{\prime})=(g\mu_B)^2\frac{i}{\hbar}
\langle[S_i (\rr,t),S_j (\rr^{\prime},t^{\prime})]
\rangle\theta(t-t^{\prime}) 
\label{kubo}
\end{equation}
Here $i=x,y$ and $S_i(\rr,t)=M_i(\rr,t)/(g\mu_B)$ are the Mn 
transverse spin-density operators.

To evaluate the correlation function (\ref{kubo}) we use the
Holstein-Primakoff boson representation of the spin operators
assuming small fluctuations around the mean-field ordered state,
$S^{+}=b\sqrt{2N_{Mn}S}$ and $ S^{-}=b^{\dagger}\sqrt{2N_{Mn}S}$
($S_x=(S^++S^-)/2$, $S_y=(S^+-S^-)/2i)$, and choosing the $\hat z$-direction
as the quantization axis. After integrating out the 
itinerant carrier degrees of freedom within the 
coherent-state path-integral formalism of the many-body problem
we obtain the partition function, 
$Z=\int {\cal D}[\bar z z] \exp (-S[\bar z z])$, with the action
given to quadratic order in $z$ and $\bar z$ 
(the complex numbers representing the bosonic degrees of freedom) by
\begin{equation}
S[\bar{z}z]={1}/{\beta V}\sum_{m,\kk}
\bar{z}(\kk,\Omega_m)(-i\Omega_m+\Pi(i\Omega_m)){z}(\kk,\Omega_m)\; .
\label{action}
\end{equation}
In Eq. (\ref{action}), 
the first term is the standard Berry's phase and the second term is 
the itinerant carrier spin polarization diagram 
for the high symmetry case where the cross terms of the form 
$\bar z \bar z$ and $z z$ vanish in $S[\bar z z]$:\cite{konig_prb01}
\begin{eqnarray}
\Pi(i\Omega)&=&\frac{N_{Mn}J_{pd}^2S}{2\beta}
\int\frac{d^3k}{(2\pi)^3}\sum_{m,a,b}
{\cal G}_a(i\omega_m,\kk) \nonumber \\
&\times&{\cal G}_b(i\omega_m+i\Omega,\kk)
\left|\langle\phi_{a}(\kk)|s^+|\phi_{b}(\kk)\rangle\right|^2 \; .
\label{pi}
\end{eqnarray}
Here ${\cal G}_a(z,\kk)$ is the single-particle
band Green's function and $\phi_{a}(\kk)$ are the band eigenstates.
From the partition function we compute directly the 
imaginary time response functions at finite Matsubara frequencies
which after their corresponding analytic continuations yield
\begin{eqnarray}
{\chi^{R}_{xx}}(\omega)&=&-(g\mu_B)^2\frac{N_{Mn}S}{2\hbar}
\frac{2\Pi^{\rm ret}(\omega)}
{\omega^2+i\delta-{\Pi^{\rm ret}}^2(\omega)}\label{sus1}
\nonumber \\
{\chi^{R}_{xy}}(\omega)&=&-i(g\mu_B)^2\frac{N_{Mn}S}{2\hbar}
\frac{2\omega}{\omega^2+i\delta-{\Pi^{\rm ret}}^2(\omega)}\; .
\label{sus2}
\end{eqnarray}
Here $\Pi^{\rm ret}$ (calculated below)
describes mathematically the retarded interaction between the Mn bosonic 
degrees of freedom due to the p-d kinetic exchange with valence band holes.

{\em Connecting classical phenomenology and microscopic theory.}
Inverting the retarded susceptibility in Eq.~(\ref{sus2})  for 
the uniform ($\kk=0$) precession mode, we obtain
\begin{eqnarray}
{\chi^{R}_{xx}}^{-1}(\omega)
&=& \frac{\hbar\Pi^{\rm ret}(\omega)}{(g\mu_B)^2N_{Mn}S}
\nonumber \\
{\chi^{R}_{xy}}^{-1}(\omega)
&=&-i\frac{\hbar\omega}{(g\mu_B)^2N_{Mn}S}\; .
\label{quantumsusc}
\end{eqnarray}
Comparing Eqs.~(\ref{quantumsusc}) and (\ref{classicalsusc}), we obtain the
microscopic contribution  to the 
Gilbert coefficient from kinetic-exchange coupling :
\begin{equation} 
\alpha=- \lim_{\omega \to 0} \Big[ {\rm Im}\Pi^{\rm ret}(\omega)/\omega \Big]
\; .
\end{equation} 
Note that ${\chi^{R}_{xy}}^{-1}$ is explicitly equal to the off-diagonal
element of $\chi^{-1}$ in Eq. \ref{classicalsusc} and 
that, also consistently, the real part of $\Pi^{\rm ret}(\omega)$, in the
limit of $\omega\rightarrow 0$, gives the magnetocrystalline contribution
to the anisotropy energy $\tilde{U}$, as explained in detail in 
Ref.~\onlinecite{konig_prb01}. 

To compute $\Pi^{\rm ret}(\omega)$, we take into account the effects of
disorder present  in the system  perturbatively by accounting  for the
finite-lifetime  of  band  quasiparticles,  which  for  simplicity  we
characterize   by  a  single   number  $\Gamma$.    The  quasiparticle
broadening  $\Gamma$  was chosen  to  be  in  the range  estimated  in
previous  detailed  studies  of    transport  properties  of  these
systems,    which    achieve    good   agreement    with    
experiment.\cite{jungwirth_0302060}   The single  particle Green's  function for
the   valence-band   quasiparticles   is   thus  written   as   ${\cal
G}_a(\kk,z)=\int_{-\infty}^{\infty}                     d\omega'/(2\pi)
A_a(\omega',\kk)/(z-\omega')$     with     a     spectral     function
$A_a(\epsilon,\kk)=\Gamma/[(\epsilon-\epsilon_{a,\kk})^2
+\Gamma^2/4]$. In the present case we take the valence-band electronic
structure  to be  described by  the six-band  Kohn-Luttinger Hamiltonian
in the presence of an effective kinetic-exchange field $h_{eff}=J_{pd}N_{Mn}
\langle S\rangle$.\cite{aniso} In a collinear ferromagnetic state and
zero temperature $\langle S\rangle = S$, and we obtain
\begin{eqnarray}
\alpha
&=&\lim_{\omega\rightarrow 0}\frac{N_{Mn}J_{pd}^2S}{4\hbar\omega}
\int\frac{d^3k}{(2\pi)^3}\sum_{a,b}
\left|\langle\phi_{a}(\kk)|s^+|\phi_{b}(\kk)\rangle\right|^2 
\nonumber\\
&&\times
\int \frac{d \epsilon}{2\pi}
A_{a,\kk}(\epsilon)A_{b,\kk}(\epsilon+\hbar\omega)
[f(\epsilon)-f(\epsilon+\hbar\omega)]\nonumber\\
&=&\frac{J_{pd}h_{eff}}{4\hbar}
\int\frac{d^3k}{(2\pi)^3}\sum_{a,b}
\left|\langle\phi_{a}(\kk)|s^+|\phi_{b}(\kk)\rangle\right|^2 
\nonumber\\
&&\times A_{a,\kk}(\epsilon_F)A_{b,\kk}(\epsilon_F)\label{alpha2}
\end{eqnarray}

\begin{figure}[h]
\includegraphics[width=3.5in]{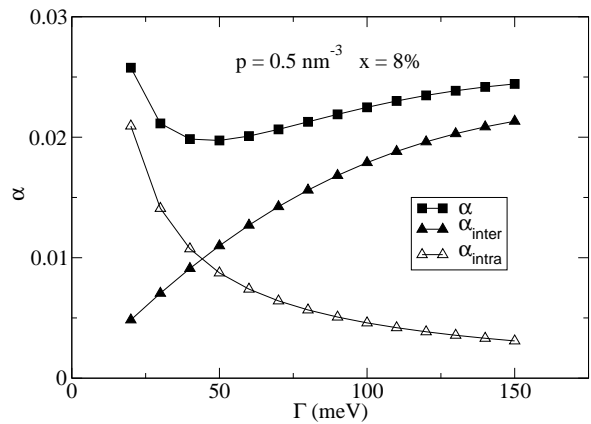}
\caption{Total Gilbert damping coefficient $\alpha$, interband contribution $\alpha_{\rm inter}$,
and intraband contribution $\alpha_{\rm intra}$ as a function quasiparticle life-time 
broadening $\Gamma$ for a 
carrier density of $0.5 {\rm nm}^{-3}$ and $x=8\%$.}
\label{gilbert_coeff2}
\end{figure}

In choosing  a spin- and  band-independent lifetime, we  are implicitly
appealing to the dominance  of spin-independent Coulomb scattering off
Mn    acceptors    and    interstitials    as    the    dominant
\cite{jungwirth_0302060}     scattering     mechanism.      Spin-orbit
interactions  enter  through their  presence  in  the intrinsic  bands
rather  than through  spin-flip quasiparticle  scattering  events.  In
this  model,   the  Gilbert  coefficient   in  Eq.~(\ref{alpha2})  has
intra-band  and  inter-band   contributions  that  have  qualitatively
different     disorder      dependences,     as     illustrated     in
Fig.~\ref{gilbert_coeff2}.   Note  that the  terminology  we use  here
recognizes that  no band  has spin-character sufficiently  definite to
justify  the  usual distinction  between  majority  and minority  spin
bands. The  intra-band term  we  refer  to  here would  be  spin-flip
scattering within a  given spin-split band in the  more usual language
and is present only because  of intrinsic spin-orbit coupling in the
host semiconductor bands.  
The  intra-band contribution to $\alpha$ is
proportional to $1/\Gamma$ at  small $\Gamma$, i.e., proportional
to the  conductivity rather than  the resistivity, and  would dominate
the  damping if disorder  were weak.   The inter-band  contribution to
$\alpha$,  on  the  other   hand,  requires  disorder  to  breach  the
wavevector separation between different  bands at the Fermi energy and
is an increasing function of  $\Gamma$.  The overall dependence of the
Gilbert  coefficient on  the  sample's mobility  is non-monotonic,  as
illustrated  in  Fig.\ref{gilbert_coeff2}, with  the  position of  the
minimum depending on  both hole and Mn-moment densities,  and on other
parameters of the DMS system.

\begin{figure}
\includegraphics[width=3.5in]{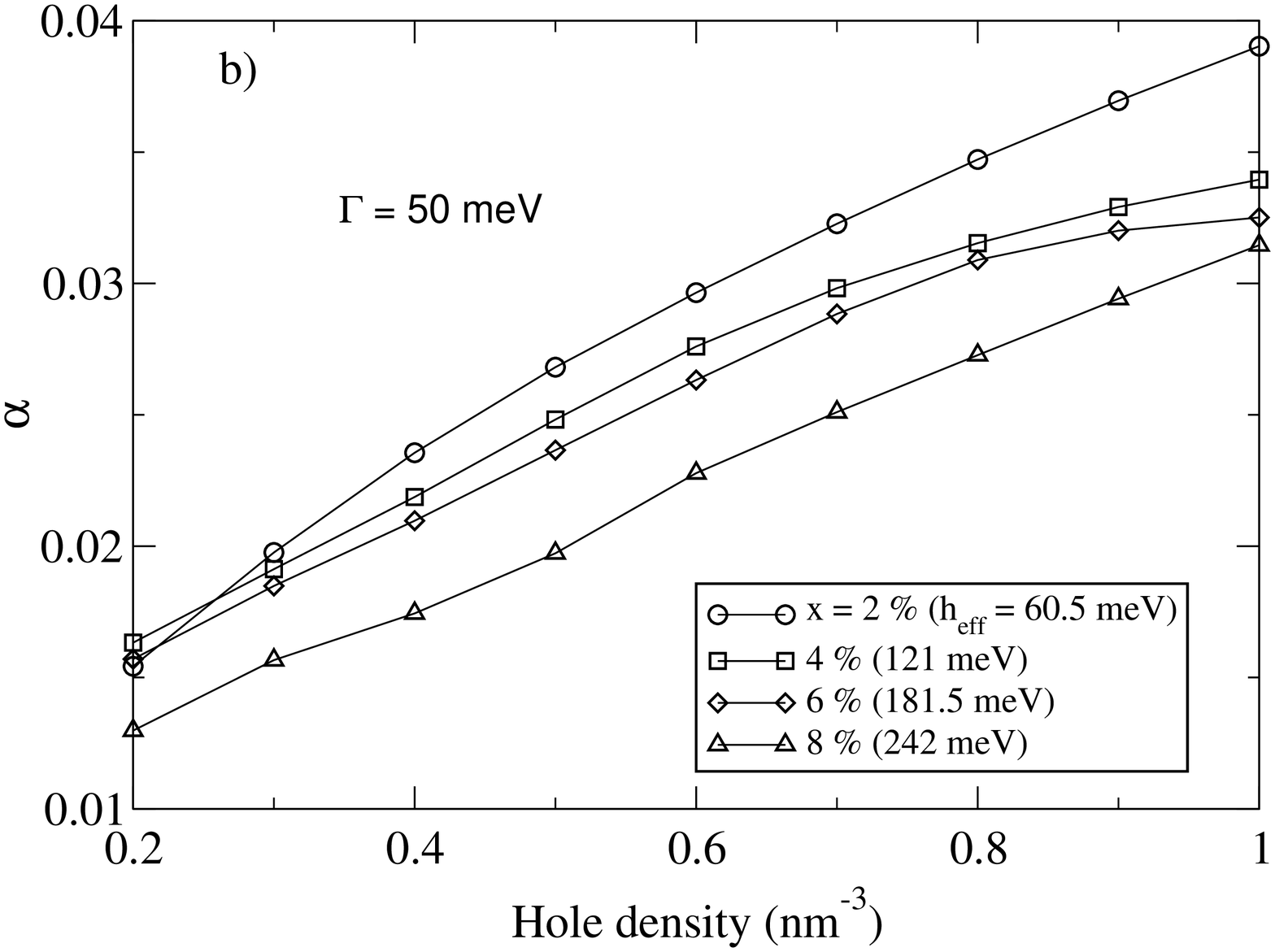}
\caption{Gilbert damping coefficient $\alpha$ as a function of
carrier density for $x=2\%$ -- 8\% and for quasiparticle life-time broadening
of $150$ meV (a) and 50 meV (b).}
\label{gilbert_coeff}
\end{figure}

The complexity and tunability of the Gilbert coefficient in DMSs is 
illustrated
in Fig.~\ref{gilbert_coeff} where we plot $\alpha$
as a function of the hole density for $\Gamma=150$ and 50 meV
and for Mn doping $x=$2 -- 8\%. 
These parameter values bracket the range typical for metallic (Ga,Mn)As DMSs.
The theory curves in Fig.~\ref{gilbert_coeff} predict that $\alpha$
increases with increasing hole density because of the higher
quasiparticle density of states at larger densities. The dependence
of  the Gilbert coefficient on $x$ is more complex. 
The prefactor $h_{eff}$ in Eq.~(\ref{alpha2})
reflects the proportionality of the kinetic-exchange coupling to
the Mn spin density and causes the Gilbert damping implied by this mechanism to
decrease with decreasing $x$ at low Mn doping. 
This behavior is clearly seen in Fig.~\ref{gilbert_coeff}(a)
for $x\le 6$~\%.
On the other hand, an opposite trend is
predicted for higher Mn-moment densities  where the effect of $h_{eff}$ on the 
intra- and inter-band matrix elements in  Eq.~(\ref{alpha2}) takes
over. In that case, larger  $h_{eff}$ values lead to  
a reduced spin mixing in the quasiparticle bands
and, therefore, to smaller Gilbert damping rates. This implicit dependence
of $\alpha$ on $h_{eff}$ is more dramatic in higher-quality samples.

We expect that the above kinetic-exchange mechanism of the Gilbert damping
will
dominate at low temperatures  where other mechanisms such as magnon-magnon
interactions vanish. 
At temperatures close to the Curie temperature, on the other hand, 
the contribution to magnetization precession 
damping due to the kinetic-exchange coupling can be ignored. The argument is based on
an approximation that combines  our zero-temperature microscopic theory of
$\alpha$ with a finite-temperature mean-field description of $h_{eff}$.\cite{aniso}
Within the mean-field model, 
$h_{eff}$ is proportional to mean Mn spin-polarization
$\langle S\rangle$ whose temperature dependence is given  
by the Brillouin function with a temperature dependent mean-field.\cite{aniso}
The curves in  Fig.~\ref{gilbert_coeff}
can therefore be approximately assigned also to a (Ga,Mn)As DMS system where
the effective field value changes through the temperature-dependent 
average Mn-spin polarization
rather than through
the Mn-doping parameter $x$. Large values of $h_{eff}$ correspond to low temperatures
in this picture and, as seen from Fig.~\ref{gilbert_coeff}, the temperature-dependence
of $\alpha$ in this regime is quite complex and sensitive to  details of the DMS sample structure.
Generally, Fig.~\ref{gilbert_coeff} suggests an initial increase of $\alpha$ with increasing
temperature in samples with a large density of Mn moments, a nearly constant $\alpha$ for
intermediate doping, and a suppression of $\alpha$ in samples with 
low Mn content. At high temperatures (small $h_{eff}$),
the kinetic-exchange-driven Gilbert damping rate will gradually decrease towards zero
with increasing temperature.    

\begin{figure}
\includegraphics[width=3.5in]{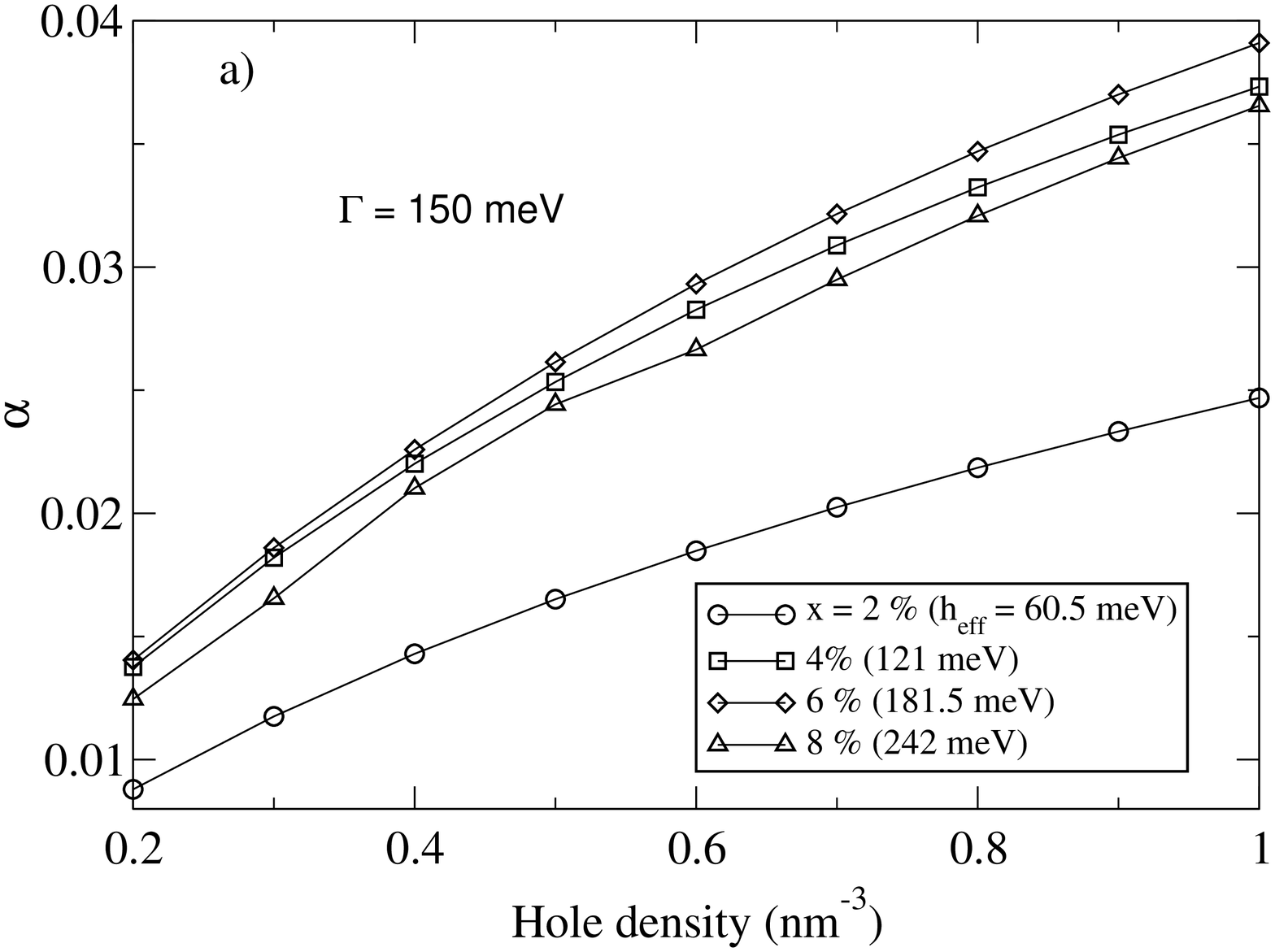}
\end{figure}

\section{Experimental FMR linewidth}
\begin{figure}
\includegraphics[width=3.5in]{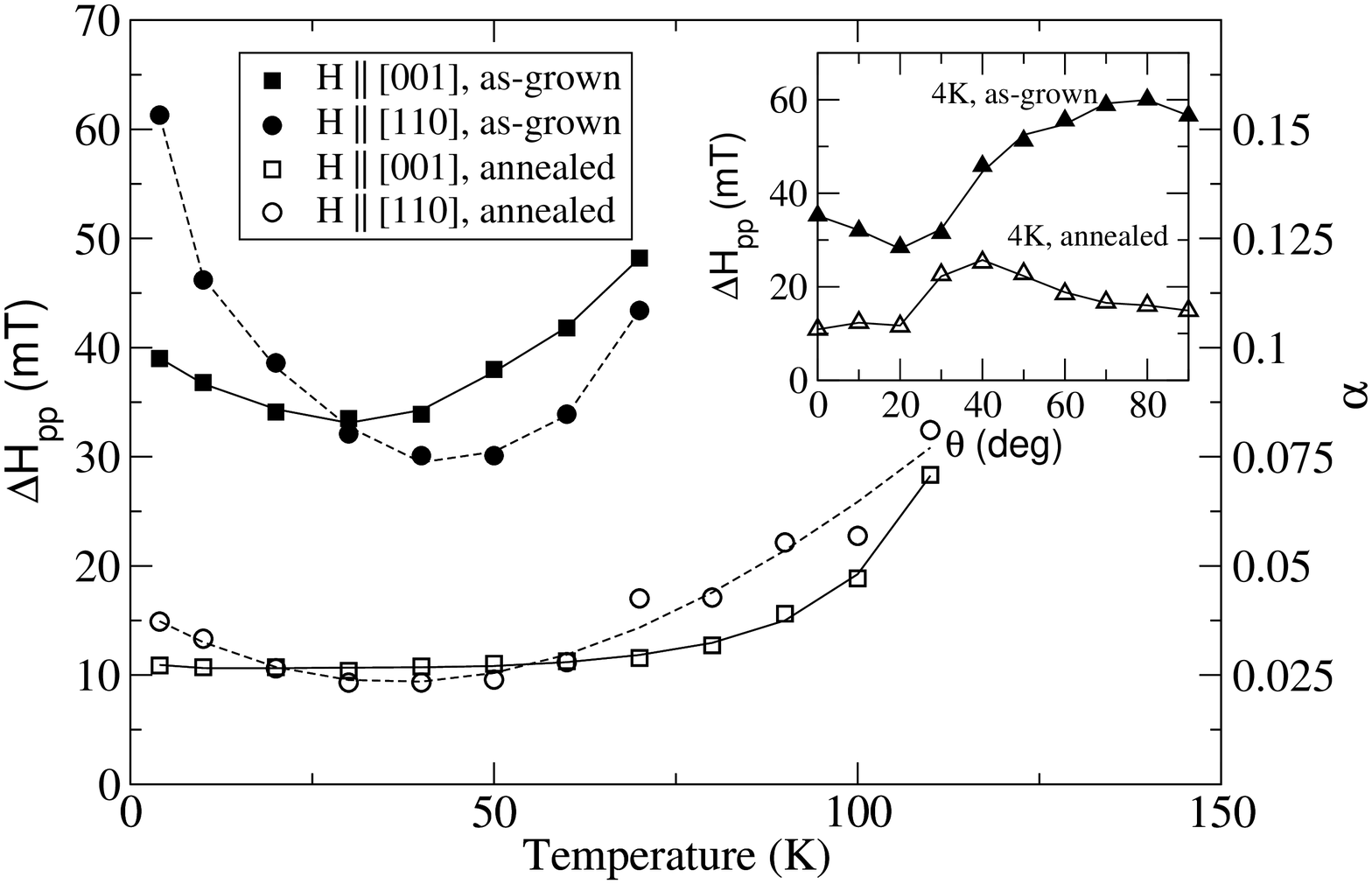}
\caption{Experimental peak-to-peak FMR linewidth in as-grown
(filled symbols) and annealed (open symbols) Ga$_{0.92}$Mn$_{0.08}$As
samples measured as a function of temperature for [001] and [110] 
dc magnetic field orientations (main plot) and as a function
of the field angle at 4~Kelvin (inset).}
\label{fmr}
\end{figure}

We now discuss our experimental FMR linewidth data recorded for 
the 120~nm thin Ga$_{1-x}$Mn$_x$As layer
with $x=8$~\% grown on GaAs (001) substrate. 
The FMR measurements were carried out at 9.46~GHz, with the external dc
magnetic field applied at different angles $\theta$ with respect to the growth
axis ($\theta =0$ corresponds to the [001] crystal direction and
$\theta =90^{\circ}$ to the [110] direction).
The Mn concentration in the sample was estimated from x-ray
diffraction measurement of the lattice constant.  The critical temperature
in the as-grown ($T_c=65$~K) and annealed ($T_c=110$~K) samples
were determined from SQUID magnetization measurements.
A more detailed description of the sample properties and 
of our experimental set-up can be found 
elsewhere.\cite{furdyna} To analyze the measured peak-to-peak FMR linewidth
$\Delta H_{pp}$, plotted
in Fig.~\ref{fmr}, we recall the following general relation between
$\Delta H_{pp}$ and $\alpha$:\cite{vonsovskii,frait}
\begin{equation}
\Delta H_{pp}(\omega)= \Delta H_{pp}(0) + \frac{2}{\sqrt{3}}\frac{\omega}
{g\mu_B} \alpha \; .
\label{dhpp}
\end{equation} 
Here $\Delta H_{pp}(0)$ describes broadening due to sample
inhomogeneity which is assumed to be frequency independent 
\cite{frait,platow} but can depend on the dc field orientation.
The second term in Eq.~(\ref{dhpp}) arises from the Gilbert damping term in the 
LLG equation~(\ref{llg}), which gives a contribution to the 
FMR linewidth which is linearly proportional to $\omega$
and independent of the static magnetic field direction, if the magnetization
is aligned with the field\cite{platow} (and the dependence of $\alpha$ on
the magnetization orientation can be neglected, as mentioned in the
previous section\cite{rem_alpha_angle}).  
This condition is satisfied in our sample
since the FMR resonance field is larger than the magnetic field at which
saturation magnetizations for different field orientations coincide.\cite{furdyna} 

The strong dependence of the FMR linewidth on the field
orientation in the as-grown sample suggests that magnetic inhomogeneities
in the ferromagnetic layer contribute significantly to the FMR broadening.
As seen in both the main panel and the  inset of Fig.~\ref{fmr}, the angle dependence
of $\Delta H_{pp}$ becomes weaker in the annealed sample and also the
overall magnitude of $\Delta H_{pp}$ is conspicuously reduced. 
This observation
is consistent with the improved quality of the sample (as indicated e.g. by
the enhanced $T_c$) and implies that the leading contribution
to the FMR linewidth might in this case 
come from the homogeneous (Gilbert damping)
broadening. 
The right y-axis in the main plot of  
Fig.~\ref{fmr} represents the experimental
Gilbert coefficient obtained from the measured $\Delta H_{pp}$
and from Eq.~(\ref{dhpp}), assuming $\Delta H_{pp}(0)=0$. Experimental
low-temperature values of $\alpha$ in the annealed sample
are around 0.03. As seen from Fig.~\ref{gilbert_coeff}, these values
of the Gilbert coefficient can be fully explained by the p-d kinetic-exchange
mechanism of the damping of magnetization precession. However, because 
the density of
Mn ions and their distribution in the lattice as well as
the density of holes 
are not precisely known in our sample, a fully quantitative comparison between
theory and experiment is not possible.  The results suggest that 
experimental studies of the {\em frequency-dependent} FMR linewidth 
in high quality DMS samples would be very valuable for
understanding the complex behavior of the Gilbert damping coefficient
predicted in the theoretical part of this paper, and 
in separating those effects 
that are arising from the
inhomogeneity within the epitaxially grown thin films.  
 
\section{Current induced magnetization switching}
Using theoretical values for the Gilbert coefficient and anisotropy 
energy, \cite{aniso} 
we now estimate the critical current for the spin-transfer induced 
magnetization switching in a ferromagnetic semiconductor multilayer
structure. In general, spin-polarized perpendicular-to-plane currents 
in magnetic multilayer systems can transfer spin between 
magnetic layers and exert current-dependent torques.
For a trilayer
structure, arguments based on the conservation of the
angular momentum suggest the following form  
of the torques on the two magnetic layers:\cite{slonczewski_jmmm96}
\begin{equation}
\tau_{1(2)}\propto \frac{I_s}{eMV}\hat{\MM}_{1(2)}\times
(\hat{\MM}_1\times\hat{\MM}_2) \; ,
\label{torque}
\end{equation}
where $\hat{\MM}_{1(2)}=\MM_{1(2)}/M$ 
and $I_s$ is the spin-polarized electric current. 
The sign of the torque depends
on the sign of the current, so that magnetization vectors in the two magnetic 
layers can be aligned parallel or anti-parallel by current flowing in one or the opposite
direction. In a spin-valve 
structure with one ``hard'' and one ``soft'' magnetic layer, switching occurs when
the torque in the soft layer overcomes the damping and the anisotropy terms. 

There have been a series of theoretical papers \cite{sttheory}
aimed at the quantitative 
description of spin currents and their effects on magnetization switching
in metallic spin-valve structures. The theories are based on a two-channel
model (spin-up and spin-down) and account for spin accumulation
effects in the magnetic multilayers and spin transfer effects
due to reflection at the ferromagnet/normal layer interface and due
to the averaging mechanism associated with rapid precession of electron spins
after entering the ferromagnet. The two-channel model is not applicable for
semiconductor valence bands with strong spin-orbit coupling, complicating the 
quantitative description of spin currents in DMSs. Strong spin-orbit
coupling leads to a reduced 
spin-coherence time. However, the exchange coupling between the Mn-moments 
and hole carriers will make this time larger in 
ferromagnetic than in non-magnetic p-type semiconductors.
Experimentally, magnetic
information can be transported by charge carriers in DMS multilayers,
despite strong spin-orbit coupling, 
as demonstrated, e.g., by the observation of the 
giant magnetoresistance effects.\cite{tanaka_prl01}

For an order-of-magnitude estimate of the switching current in a 
(Ga,Mn)As-based magnetic trilayer structure, 
we approximate the spin-current as
$I_s\approx I\langle s\rangle$, where $I$ is the electric current and
$\langle s\rangle$ is the mean-field spin polarization of the itinerant
holes in the (Ga,Mn)As layers.\cite{aniso} 
Adding the torque term (\ref{torque})
to Eq.~(\ref{llg}) for the soft magnetic layer
we obtain an effective damping rate 
$D=(\tilde{U}\alpha-\tilde{I})/(1+\alpha^2)$, where
$\tilde{I}=I\langle s\rangle/(eN_{Mn}SV)$. An instability occurs
at $D=0$ and the corresponding critical current density for
the magnetization switching 
is then given by
$j_c={eU\alpha d}/{\hbar\langle s\rangle} \; $.
Assuming a thickness $d\sim10$~nm for the soft ferromagnetic layer and
typical parameters of a Ga$_{0.95}$Mn$_{0.05}$As DMS, 
$U\sim 1$~kJ~m$^{-3}$, $\alpha\approx 
0.02$, and $\langle s\rangle\approx 0.3$, the critical current
$j_c\sim 10^5$~A~cm$^{-2}$. This estimate 
is two orders of magnitude smaller than critical
switching currents characteristic of  
metallic spin-valves,\cite{stexp,sttheory} 
primarily due to smaller saturation moments and 
anisotropy energies in the DMSs.  
Since the resistivities are only 2 to 3 orders of
magnitude larger in DMSs than in metals, observation of this effect should 
be experimentally feasible in a ferromagnet/normal/ferromagnet
semiconductor spin-valve structure.\cite{hideo}

We note that 
achieving low critical currents is particularly important for magnetic
{\em tunnel} junctions that are used in non-volatile memory devices. To avoid
self-field effects, that lead to a spin vortex state rather than to a 
complete reversal and therefore to a smaller giant
magneto-resistance effect,\cite{vortex} the in-plane 
diameter $r$ of  metallic
spin-valve devices with critical current densities $j\sim 10^7$~A~cm$^{-2}$
must be of order $\sim 100$~nm.\cite{stexp}  
Magnetic tunnel junctions have
resistances too high for applications when patterned to such small sizes.
Since the Oersted field scales as $\sim rj$ and the critical currents for
spin-torque-induced switching we predict are 
two orders of magnitude smaller in
DMSs than in metals, 1-10~$\mu$m size semiconductor tunnel junctions
might still show sufficiently
weak self-field effects and, therefore,
a complete current-induced switching.

\section{Summary}
In this article we have studied magnetization precession damping in 
ferromagnetic semiconductor (Ga,Mn)As alloys. 
We have attempted to employ  theoretical analysis 
 combined with existing experimental information
to obtain the Gilbert damping coefficient and 
to predict the scale of critical currents for spin-transfer magnetization-reversal in these systems.
In spin-transfer induced reversal, damping of
magnetization precession  competes with current-induced 
spin torques and determines the scale of the current required to achieve reversal.

Our theoretical analysis examines the mechanism that we expect to dominate 
at low temperatures in high-quality samples,
due to the coupling of the d-level local moments to 
valence band holes.  We derive an explicit expression for the Gilbert coefficient conventionally
used to characterize damping in experimental studies, by comparing microscopic linear
response theory with the linear response limit of the phenomenological Landau-Lifshitz-Gilbert
equations, and study how the values predicted by this model for the Gilbert damping coefficient
depend on the hole density and on the size of the mean-field exchange interaction experienced
by valence-band holes in the ferromagnetic state.  We find that the magnitude predicted
for the Gilbert coefficient, $\sim 0.03$, is consistent with experiment, 
but that the observed dependence
on the 
external field and magnetization orientation is larger than can be accounted for by this   
mechanism.  The experimental FMR linewidth appears to have an inhomogeneous broadening 
contribution that is not included in our theoretical modeling
developed for homogeneous bulk systems.
The uncertainty that presently exists in the relative importance of these two broadening mechanisms
could be reduced by frequency-dependent FMR studies.

In our view, the portion of the FMR linewidth broadening that is due to inhomogeneity should not,
to a first approximation, be included in assessing the competition between spin-torques and 
spin-precession damping.  We have therefore used our theoretical value
for damping in a homogeneous system  
to estimate the critical currents required
for achieving magnetization reversal and obtained 
$j_c\sim 10^{5} {\rm A {cm}^{-2}}$.  

\begin{acknowledgments}
The authors acknowledge helpful discussions with T. Dietl, Z. Frait, 
and H. Ohno.
This work was supported in part by the Welch Foundation, 
the DOE under grant DE-FG03-02ER45958,
the Grant Agency of the Czech Republic under grant 202/02/0912
the Research Corporation under grant CC5543 the DARPA SpinS Program 
and by the NSF-NIRT under grant DMR-0210519.
\end{acknowledgments}



\begin{references}
\bibitem{kittel_pr56}
C. Kittel and A.H. Mitchell, Phys. Rev. {\bf 101}, 1611 (1956);
A.H. Mitchell, {\em ibid} {\bf 105}, 1439 (1957).

\bibitem{dietl_handbook94}
T. Dietl in{\em Handbook on Semiconductors} (Elsevier, 
Amsterdam, 1994).

\bibitem{ohno_jmmm99}
H. Ohno, J. Magn. Magn. Mater, {\bf 200}, 110 (1999);

\bibitem{mm03}
T.C. Schulthess, Bull. Am. Phys. Soc. 2003, March Meeting, K30-1;
P.H. Dederichs, K. Sato, H. Katayama-Yoshida,J. Kudrnovsk\'y, {\em ibid},
S24-5.

\bibitem{bookchapter}
J. K\"{o}nig, J. Schliemann, T. Jungwirth,
and A.H. MacDonald,
in
{\em Electronic Structure and Magnetism of Complex Materials},
edited by D.J. Singh and D.A. Papaconstantopoulos (Springer Verlag, Berlin
2003).
\bibitem{dietl_sci00}
T. Dietl, H. Ohno, F. Matsukura, J. Cibert, and
D. Ferrand, Science {\bf 287}, 1019 (2000).

\bibitem{tc}
T. Jungwirth, J. K\"{o}nig, J. Sinova, J. Ku\v{c}era, and
A.H. MacDonald, Phys. Rev. B {\bf 66}, 012402 (2002);
K. W. Edmonds, K. Y. Wang, R. P. Campion, A. C. Neumann, 
C. T. Foxon,B. L. Gallagher, and P. C. Main,  Appl. Phys. Lett 
{\bf 81}, 3010 (2002).


\bibitem{jungwirth_0302060}
T. Jungwirth, Jairo Sinova, K.Y. Wang, K. W. Edmonds, 
R.P. Campion, B.L. Gallagher, C.T. Foxon, Q. Niu, A.H. MacDonald,
Appl. Phys. Lett. 83, 320 (2003).


\bibitem{sinova_prb02}
Jairo Sinova, T. Jungwirth, S.- R. Eric Yang, J. Ku\v{c}era, 
and A.H. MacDonald,  
Phys. Rev. B {\bf 66}, 041202 (2002).

\bibitem{aniso}
T. Dietl, H. Ohno, F. Matsukura,  Phys. Rev. B {\bf 63}, 195205 (2001);
M. Abolfath, T. Jungwirth, J. Brum, and A.H. MacDonald, 
{\em ibid} {\bf 63}, 054418 (2001).

\bibitem{konig_prb01}
J. K\"{o}nig, T. Jungwirth, and A.H. MacDonald,  
Phys. Rev. B {\bf 64}, 184423 (2001).

\bibitem{stiffness}
S.T.B. Goennenwein, T. Graf, T. Wassner, M.S. Brandt, 
M. Stutzmann, J. B. Philipp, R. Gross, M. Krieger, K. Z\"{u}rn, 
P. Ziemann, A. Koeder, S. Frank, W. Schoch, and A. Waag,  
Appl. Phys. Lett. {\bf 82}, 730 (2003).

\bibitem{domains}
T. Shono,T. Hasegawa, T. Fukumura, F. Matsukura and H. Ohno, 
Appl. Phys. Lett. {\bf 77}, 1363 (2000);
T. Dietl, J. K\"{o}nig, and A. H. MacDonald,  
Phys. Rev. B {\bf 64}, 241201 (2001).

\bibitem{furdyna} X. Liu, Y. Sasaki, and J.F. Furdyna, Phys. Rev. B.
{\bf 67}, 205204 (2003).

\bibitem{slonczewski_jmmm96}
J.C. Slonczewski, J. Magn. Magn. Mater. {\bf 159}, L1 (1996);
L. Berger, Phys. Rev. B {\bf 54}, 9353 (1996).

\bibitem{stexp}
E.B. Myers, D.C. Ralph, J.A. Katine, R.N. Louie, and R.A. Buhrman,
Science {\bf 285}, 867 (1999); M. Tsoi, A.G.M. Jansen, J. Bass, W.-C. Chiang,
V. Tsoi, and P. Wyder, Nature {\bf 409}, 46 (2000); W. Weber, S. Riesen,
H.C. Siegmann, Science {\bf 291}, 1015 (2001); F.J. Albert,
N.C. Emley, E.B. Myers, D.C. Ralph, and R.A. Buhrman, Phys. Rev. Lett.
{\bf 89}, 226802 (2002).

\bibitem{sttheory}
Ya. B. Bazaliy, B.A. Jones, and
Shou-Cheng Zhang, Phys. Rev B {\bf 57}, R3213 (1998);
J.C. Slonczewski, J. Magn. Magn. Mater. {\bf 195}, L261 (1999);
J.C. Slonczewski, cond-mat/0208207; M.D. Stiles and A. Zangwill,
J. Appl. Phys. {\bf 91}, 6812 (2002); Phys. Rev. B {\bf 66}, 014407
(2002); S. Zhang, P.M. Levy, and A. Fert, Phys. Rev. Lett. {\bf 88},
236601 (2002).

\bibitem{field_tc}
B.H. Lee, T. Jungwirth, and A.H. MacDonald, Phys. Rev. B {\bf 61}, 15606 (2000);
H. Ohno, D. Chiba, F. Matsukura, T. Omiya, E. Abe, T. Dietl, Y. Ohno, and K. Ohtani, 
Nature {\bf 408}, 944 (2000).

\bibitem{field_reversal}
Byounghak Lee, T. Jungwirth, and A.H. MacDonald, Phys. Rev. B {\bf 65}, 193311 (2002);
D. Chiba,1 M. Yamanouchi,1 F. Matsukura, H. Ohno,  Science {\bf 301}, 943 (2003).

\bibitem{vortex}
Yaowen Liu, Zongzhi Zhang, P. P. Freitas, and J. L. Martins,
 Appl. Phys. Lett. {\bf 82}, 2871 (2003).

\bibitem{rem_alpha_angle} Stronger angle dependence of $\alpha$ is expected only in
samples with large lattice-matching strains induced, e.g.,  by choosing a different
III-V semiconductor as a substrate.

\bibitem{vonsovskii}
S. V. Vonsovskii, {\em Ferromagnetic Resonance} (Pergamon, Oxford, 1966).

\bibitem{frait}
F. Schreiber, J. Pflaum, Z. Frait, Th. M\"{u}hge, and J. Pelzl,
Solid State Commun. {\bf 93}, 965 (1995).

\bibitem{platow}
W. Platow, A.N. Anisimov, G.L. Dunifer, M. Farle, and K. Baberschke,
Phys. Rev. B {\bf 58}, 5611 (1998).

\bibitem{tanaka_prl01}
M. Tanaka and Y. Higo,
Phys. Rev. Lett. {\bf 87}, 026602 (2001).

\bibitem{hideo}
H. Ohno, private commun.


\end{references}
\end{document}